\documentclass{epl2}
\usepackage{amssymb}
\usepackage{amsmath}

\shorttitle{Title} 
\shortauthor{F. Author \etal}
\institute{                    
  \inst{1} Condensed Matter Theory Group, CPMOH, Universit\'{e} de Bordeaux and CNRS. F-33405 Talence, France\\
  \inst{2} Also at Institut Universitaire de France. 
}
\abstract{
We study the influence of the configuration of the majority and minority spin subbands of electron spectra on the properties of atomic-scaled superconductor-ferromagnet S-F-S and F-S-F hybrid structures. At low temperatures, the S/F/S junction is either a 0 or  junction depending on the energy shift between S and F materials and the anisotropy of the Fermi surfaces. We found that the spin switch effect in F/S/F system can be reversed if the minority spin electron spectra in F metal is of the hole-like type.}
\input{tcilatex}

\begin{document}

\title{Proximity effect in atomic-scaled hybrid superconductor/ferromagnet
structures: crucial role of electron spectra.}
\author{X. M\textsc{ontiel}\inst{1} \and D. G\textsc{usakova}\inst{1} \and %
M. D\textsc{aumens}\inst{1} \and A. B\textsc{uzdin}\inst{1,2}}
\maketitle

\textbf{1. Introduction}\textit{-} Phenomena arising in the
superconductor-ferromagnet (S/F) hetero-structures attract a growing
interest due to their potential application in spintronic and quantum
computation devices [1-3]. For example, the so-called spin-switch effect
(SSE) (also called spin-valve effect) occurs when the critical temperature
of the superconductor in the F/S/F structure depends on the mutual
orientation of the ferromagnetic layers magnetizations. Moreover the S/F/S $%
\pi -$junctions with the changing superconducting order parameter phase
shift are good candidates for the quantum computer elements, q-bit \cite%
{Yamashita_2005}.

Critical current oscillations as a function of the exchange field and
thickness of the ferromagnetic spacer in a S/F/S Josephson junctions have
been predicted \cite{Bulaevskii_19822} and associated with the $0-\pi $
phase transition. Experimentally the evidence of such $0-\pi $ phase
transition has been obtained first for the nanoscaled Josephson junctions
S/F/S with a weak ferromagnetic interlayers in Cu$_{x}$Ni$_{1-x}$ \cite%
{Ryazanov_Aarts_2001} and PdNi alloys \cite{Kontos_2002}. The $0-\pi $ phase
transition was also predicted in layered compounds within a simple model of
alternating F and S atomic layers [8-9]. This prediction is relevant for
ruthenocuprates RuSr$_{2}$GdCu$_{2}$O$_{8}$ which are natural S/F layered
compounds \cite{Mac_Laughlin_1999}.

The F/S/F trilayer also exhibits interesting spin dependent phenomena. It
was predicted [11-13] and observed [14-19] that the critical temperature is
higher for antiparallel (AP) magnetization configuration than for the
parallel (P) one. Besides this so called normal SSE, several recent
experiments report on an inverse SSE whereby the parallel configuration is
more favorable for the superconductivity than the antiparallel one [20-25].

The situation with inverse SSE investigated in \cite{Aarts_J_2006} is
somewhere controversial because in the similar Py/Nb/Py system in \cite%
{O_Birge_2006_2} the normal SSE was observed. The possible explanation of
the inverse SSE observed in \cite{Aarts_J_2006} has been recentlty proposed
in \cite{Zhu_2009} and it is related with a stray magnetic field generated
in the AP configuration. The similar arguments may explain inverse SSE
observed in [22,23]. However this reasoning do not work for the inverse SSE
revealed in \cite{A_Singh_2007} where the magnetizations were perpendicular
to the layers. therefoe due to the demagnetization the influence of orbital
effect was excluded.

In this letter, we investigate the $0-\pi $ phase transition and SSE for
more realistic band structure. In particular, the $\pi -$phase might be
suppressed when the ferromagnet bands are sufficiently energy-shifted with
respect to the superconductor ones. The anisotropy of the quasi particle
spectra is also of primary importance for the $0-\pi $ phase transition.
This could explain why the $\pi -$phase was not observed in Ruthenocuprates 
\cite{Nachtrab_2006}. Furthermore, if the majority and minority spin
subbands have opposite electron/hole character, we predict the inverse SSE
providing a possible explanation of results \cite{A_Singh_2007}. 
\begin{figure}[tbp]
\begin{center}
\includegraphics[scale=0.3,angle=0]{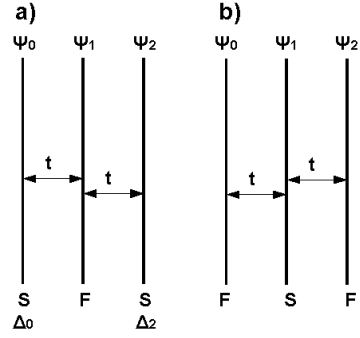}
\end{center}
\caption{Geometry of the three-layer system. Here $\protect\psi _{0}$,$~%
\protect\psi _{1}$ and $\protect\psi _{2}$ are the electron annihilation
operators in the corresponding layer.}
\label{FIG_SFSFSF}
\end{figure}

\textbf{2. Model}\textit{\ -} We start with an exactly solvable model \cite%
{Andreev_Buzdin_OsgoodIII_1991} of alternating superconducting and
ferromagnetic atomic metallic layers. The electron's motion is described in
the F layers by the spin-dependent energy spectrum $\xi _{\sigma }\left( 
\mathbf{k}\right) $ and by spin independent energy spectrum $\xi _{s}\left( 
\mathbf{k}\right) $ in the S layers. Three basic parameters characterize the
system : $t$ is the transfer energy between the F and S layers, $\lambda $
is the Cooper pairing constant which is assumed to be nonzero in S layers
only, and $h$ is the constant exchange field in the F layers only. It is
supposed \ that the coupling between the layers is realized via the transfer
integral $t$, which is relatively small $\left( t\ll T_{c}\right) $, so the
superconductivity can coexist with ferromagnetism in the adjacent layers.
The Hamiltonian of the system can be written as $H=H_{0}+H_{BCS}+H_{t}$ with:%
\begin{align}
& H_{0}=\tsum_{n,\sigma ,\mathbf{k}}\left[ \xi _{\sigma }^{n}\left( \mathbf{k%
}\right) \psi _{\sigma ,n}^{+}\left( \mathbf{k}\right) \psi _{\sigma
,n}\left( \mathbf{k}\right) \right] ,  \label{hamiltonien_generale} \\
& H_{BCS}=\tsum_{n,\mathbf{k}}\left[ \Delta _{n}^{\ast }\psi _{\downarrow
,n}^{+}\left( \mathbf{k}\right) \psi _{\uparrow ,n}^{+}\left( -\mathbf{k}%
\right) +\Delta _{n}\psi _{\uparrow ,n}\left( \mathbf{k}\right) \psi
_{\downarrow ,n}\left( -\mathbf{k}\right) \right] ,
\label{hamiltonien_generalee} \\
& H_{t}=t\tsum_{n,\sigma ,\mathbf{k}}\left[ \psi _{\sigma ,\left( n+1\right)
}^{+}\left( \mathbf{k}\right) \psi _{\sigma ,n}\left( \mathbf{k}\right)
+\psi _{\sigma ,n}^{+}\left( \mathbf{k}\right) \psi _{\sigma ,\left(
n+1\right) }\left( \mathbf{k}\right) \right] ,  \label{hamiltonien_general}
\end{align}%
where $\psi _{\sigma ,n}^{+}\left( \mathbf{k}\right) $ is the creation
operator of an electron with spin $\sigma $ and momentum $\mathbf{k}$ in the 
$n$th layer. The BCS pairing in the S layers is treated in $H_{BCS}$ in a
mean-field approximation \cite{b.abrikosov_gorkov}. The superconducting
order parameter $\Delta _{n}$ is non zero only in the S layers. Note that
the electrons spectra \ in $\left( \ref{hamiltonien_general}\right) $ are
calculated from the Fermi energy. As usual, we introduce the normal and
anomalous Green functions \cite{b.abrikosov_gorkov} $G_{\sigma ,\sigma
^{\prime }}^{n,m}=-\left\langle T_{\tau }\left( \psi _{\sigma ,n}\left( 
\mathbf{k}\right) \psi _{\sigma ^{\prime },m}^{+}\left( \mathbf{k}\right)
\right) \right\rangle $ and $\widetilde{F}_{\sigma ,\sigma ^{\prime
}}^{n,m}=\left\langle T_{\tau }\left( \psi _{\sigma ,n}^{+}\left( \mathbf{k}%
\right) \psi _{\sigma ^{\prime },m}^{+}\left( -\mathbf{k}\right) \right)
\right\rangle $ which satisfy the system of equations:%
\begin{align}
& \left( i\omega -\xi _{\sigma }^{n}\right) G_{\sigma ,\sigma ^{\prime
}}^{n,m}-tG_{\sigma ,\sigma ^{\prime }}^{n-1,m}-tG_{\sigma ,\sigma ^{\prime
}}^{n+1,m}+\Delta _{n}^{\ast }\widetilde{F}_{-\sigma ,\sigma ^{\prime
}}^{n,m}=\delta _{nm},  \notag \\
& \left( i\omega +\xi _{-\sigma }^{n}\right) \widetilde{F}_{-\sigma ,\sigma
^{\prime }}^{n,m}+t\widetilde{F}_{-\sigma ,\sigma ^{\prime }}^{n-1,m}+t%
\widetilde{F}_{\sigma ,\sigma ^{\prime }}^{n+1,m}+\Delta _{n}G_{\sigma
,\sigma ^{\prime }}^{n,m}=0,  \label{equations_generales}
\end{align}%
where $\omega =\left( 2l+1\right) \pi T_{c}$ are the fermion Matsubara's
frequencies, $\xi _{\sigma }^{n}=\xi _{\sigma }^{n}\left( \mathbf{k}\right) $
and $n$ and $m$ the layers indices. The superconducting order parameter in
the $n$th layer satisfies the standard self-consistency equation 
\begin{equation}
\Delta _{n}^{\ast }=\left\vert \lambda \right\vert T\tsum_{\omega }\tsum_{%
\mathbf{k}}\widetilde{F}_{-\sigma ,\sigma ^{\prime }}^{n,m}.
\label{Self_consistency}
\end{equation}%
The anomalous Green function $\widetilde{F}_{-\sigma ,\sigma ^{\prime
}}^{n,m}$ of the system $\left( \ref{Self_consistency}\right) $ can be be
expressed by means of normal Green functions%
\begin{equation*}
\widetilde{F}_{-\sigma ,\sigma ^{\prime }}^{n,m}=\tsum_{p}\Delta
_{n-l}^{\ast }G_{-\sigma ,\sigma ^{\prime }}^{\omega ;n-p,m}G_{0,\sigma
,-\sigma ^{\prime }}^{-\omega ;p,m},
\end{equation*}%
where $G_{0}$ is the Green function in the absence of superconductivity
pairing, i.e. for $\Delta _{n}=0$.

\textbf{3. }$0-\pi $\textbf{\ phase transition in S/F/S system }- In this
section we examine the phase difference between the order parameters in the
adjacent superconducting layers of the S/F/S system (see figure \ref%
{FIG_SFSFSF} (a)). Here $n=0,2$ and $\xi _{s,0}=\xi _{s,2}=\xi _{s}$ for the
S layers, and $n=1$ with $\xi _{\sigma ,1}=\xi _{\sigma }$ for the F layer.
Due to the symmetry reason the order parameters \ of both S layers may
differ from each other in the phase prefactor $e^{i\varphi }$ and $%
\left\vert \Delta _{0}\right\vert =\left\vert \Delta _{2}\right\vert $ (in
fact only the situation with $\Delta _{0}=\pm \Delta _{2}$ is possible).

Solving the system of equation $\left( \ref{equations_generales}\right) $ in
the case of \ the trilayer , we find the anomalous Green functions of the S
layers. Using them, the self-consistency equation $\left( \ref%
{Self_consistency}\right) $ can be written as :%
\begin{equation}
\Delta _{0}^{\ast }=-\lambda T\left\{ \tsum_{\omega ,\mathbf{k}}\left[
G_{\uparrow \uparrow }^{\omega ;0,0}\Delta _{0}^{\ast }G_{0\downarrow
\downarrow }^{-\omega ;0,0}+G_{\uparrow \uparrow }^{\omega ;2,0}\Delta
_{2}^{\ast }G_{0\downarrow \downarrow }^{-\omega ;2,0}\right] \right\} .
\label{autoconsistence 2}
\end{equation}%
where%
\begin{align*}
& G_{0\uparrow \uparrow }^{0,0}=\tfrac{\left( a_{\uparrow }^{\ast
}-t^{2}\right) }{\left( i\omega -\xi _{s}\right) \left[ \left( a_{\uparrow
}^{\ast }-2t^{2}\right) \right] },G_{0,\uparrow \uparrow }^{2,0}=\tfrac{t^{2}%
}{\left( i\omega -\xi _{s}\right) \left[ \left( a_{\uparrow }^{\ast
}-2t^{2}\right) \right] }, \\
&
\end{align*}%
with $a_{\sigma }=\left( i\omega +\xi _{\sigma }\right) \left( i\omega +\xi
_{s}\right) $. The transition between $0$ and $\pi $ states implies a change
of the relative sign between $\Delta _{0}$ and $\Delta _{2}$. Near $T_{c}$,
taking into account that in $0-$phase, $\Delta _{0}=+\Delta _{2}$ and in the 
$\pi -$phase, $\Delta _{0}=-\Delta _{2}$, the combination of the self
consistency equations $\left( \ref{autoconsistence 2}\right) $ written in $%
0- $phase and $\pi -$phase provides one with the following expression 
\begin{equation}
\ln \left( \tfrac{T_{c}^{0}}{T_{c}^{\pi }}\right)
=2T_{c0}\tsum\limits_{\omega ,\mathbf{k}}G_{0\uparrow \uparrow }^{\omega
;2,0}G_{0\downarrow \downarrow }^{-\omega ;2,0},  \label{autoconsistence 3}
\end{equation}%
where $T_{c}^{0}$ is the superconducting critical temperature when the
system is in the $0-$phase, $T_{c}^{\pi }$ is the superconducting critical
temperature when the system is in the $\pi -$phase and $T_{c0}$ is the bare
mean-field critical temperature of the single superconductivity layer. The
logarithm can be simplified as $\ln \left( T_{c}^{0}/T_{c}^{\pi }\right)
\simeq \left( T_{c}^{0}-T_{c}^{\pi }\right) /T_{c}^{0}=\Delta T/T_{c}^{0}$
because the critical temperature variation is small $\left( \left(
T_{c}^{0}-T_{c}^{\pi }\right) /T_{c}^{0}\ll 1\right) $.

The sign of the right hand side in $\left( \ref{autoconsistence 3}\right) $
determines whether it is $0$ or $\pi-$phase which is realized. Indeed, the
transition occurs in the state with higher critical temperature. We are
interested in the situation when the right hand side in\ $\left( \ref%
{autoconsistence 3}\right) $ is negative, i.e. $T_{c}^{\pi}>T_{c}^{0}$ and
we deal with the $\pi-$phase difference of order parameters between the two
superconducting layers.

First, we concentrate on the case with isotropic electron's dispersion when
Fermi surfaces are circular. The calculation of the critical temperature, in
the case of isotropic Fermi surface and symmetrical ferromagnet energy band
splitting $\xi ^{\uparrow (\downarrow )}=\xi _{s}\mp h$ has been performed
in \cite{houzet_buzdin.2001_2}. Here we address the more general case with $%
\xi _{\uparrow }=\xi _{s}+E^{\uparrow }$ and $\xi _{\downarrow }=\xi
_{s}+E^{\downarrow }$ where $E^{\uparrow \left( \downarrow \right) }$ is the
energy difference of the electron energy spectrum with spin up (down)
compared to the superconductor energy spectrum $\xi _{s}$. The calculation
of the sum over momentum $\mathbf{k}$ in $\left( \ref{autoconsistence 3}%
\right) $ may be transformed into an integration over $\xi _{s}$, i.e. $%
\sum_{k}\longrightarrow N\left( 0\right) \int \int d\xi _{s}d\theta $, where 
$N\left( 0\right) $ is the density of state at the Fermi energy $E_{F}$.
Performing the integration over energy $\xi _{s}$ we confronted with several
possible situations according to the values and signs of $E^{\uparrow }$ and 
$E^{\downarrow }$ that may be classified as follows : 1) $E^{\uparrow }\gg
T_{c0}$, but $E^{\downarrow }$ is of the order of $T_{c0}$, 2) $E^{\uparrow
}\gg T_{c0}$, $E^{\downarrow }\gg T_{c0}$ and 3) $E^{\uparrow }\gg T_{c0}$, $%
E^{\downarrow }\ll T_{c0}$.

In the case 1) when $E^{\uparrow }\gg T_{c0}$ and $E^{\downarrow }$ is of
the order of $T_{c0}$, the electrons up \ energy band is strongly shifted in
comparison to the superconductor one and electrons down have energy band
close to the superconductor one. The critical temperature difference in the
limit $t\ll T_{c}$ ,with $\omega >0$, becomes :%
\begin{equation}
\tfrac{\Delta T}{T_{c}^{0}}=\tfrac{7\pi t^{4}\zeta \left( 3\right) }{%
8E^{\uparrow }E^{\downarrow }\pi ^{3}T_{c0}^{2}}-\tfrac{t^{4}}{8E^{\uparrow
}E^{\downarrow ^{2}}\pi T_{c0}}\func{Im}\left( \Psi ^{\prime }\left( \tfrac{1%
}{2}-\tfrac{iE^{\downarrow }}{4\pi T_{c0}}\right) \right) .  \label{I_0}
\end{equation}%
where $\Psi ^{\prime }$ is the first derivative digamma function. From $%
\left( \ref{I_0}\right) $ we deduce that $\Delta T=T_{c}^{0}-T_{c}^{\pi }>0$
and the S/F/S system is in the $0-$phase if $E^{\uparrow }$ and $%
E^{\downarrow }$ have both the same sign. However, $\Delta T<0$ and the
S/F/S system is in the $\pi -$phase if $E^{\uparrow }$ and $E^{\downarrow }$
have opposite sign \textit{e.g}. $E^{\uparrow }$ positive and $E^{\downarrow
}$ negative. Thus, the existence of the $\pi -$phase depends on the energy
band shift between the ferromagnetic spectra and the superconductors spectra.

If we switch the role of $E^{\downarrow}$ and $E^{\uparrow}$ in $\left( \ref%
{autoconsistence 3}\right) $ i.e. $E^{\downarrow}\gg T_{c0}$ and $%
E^{\uparrow}$ is of the order of $T_{c0}$, the conclusions are similar. When 
$E^{\uparrow}$ and $E^{\downarrow}$ have both the same sign, the S/F/S
system is in the $0-$phase. When $E^{\uparrow}$ and $E^{\downarrow}$ have
opposite sign, the S/F/S system is in the $\pi-$phase.

In the case 2), when $E^{\uparrow }\gg T_{c0}$ and $E^{\downarrow }$ $\gg
T_{c0}$, the electrons up and down\ have both a strong energy band shift
compared to the superconducting energy spectrum. The critical temperature
difference becomes $\Delta T/T_{c}^{0}=7\pi t^{4}\zeta \left( 3\right)
/\left( 8E^{\uparrow }E^{\downarrow }\pi ^{3}T_{c0}^{2}\right) $. The $\pi -$%
phase only appears if $E^{\uparrow }$ and $E^{\downarrow }$ have an opposite
sign.

In the case 3) when $E^{\uparrow }\gg T_{c0}$ and $E^{\downarrow }$ $\ll $ $%
T_{c0}$, the electrons with spin up have a strong energy band shift compared
to the superconducting energy spectrum and the electrons with spin down a
very small one. The critical temperature difference becomes $\Delta
T/T_{c}^{0}=31E^{\downarrow }t^{4}\zeta \left( 5\right) /\left( 64\pi
^{4}T_{c0}^{4}E^{\uparrow }\right) $. Again the $\pi -$phase appears if $%
E^{\uparrow }$ and $E^{\downarrow }$ have an opposite sign. Here, we note
that $\Delta T$ depends linearly on $E^{\downarrow }$. As a consequence,
when $E^{\downarrow }$ goes to zero (superconductor and electrons down
ferromagnet energy band coincides) then $T_{c}^{0}=T_{c}^{\pi }$ in our
approximation and to determine the type of the ground state it is needed to
calculate $\Delta T$ at higher order over $t$ approximation.

As a conclusion, the $\pi -$phase does not exist if the energy band edges
for both spin orientation of the ferromagnet are higher or lower than that
of superconductor one. Consequently, the energy shift between the
ferromagnet electron band and the superconductor electron band strongly
influences the $\pi -$phase existence. 
\begin{figure}[tbp]
\begin{center}
\includegraphics[scale=0.15,angle=0]{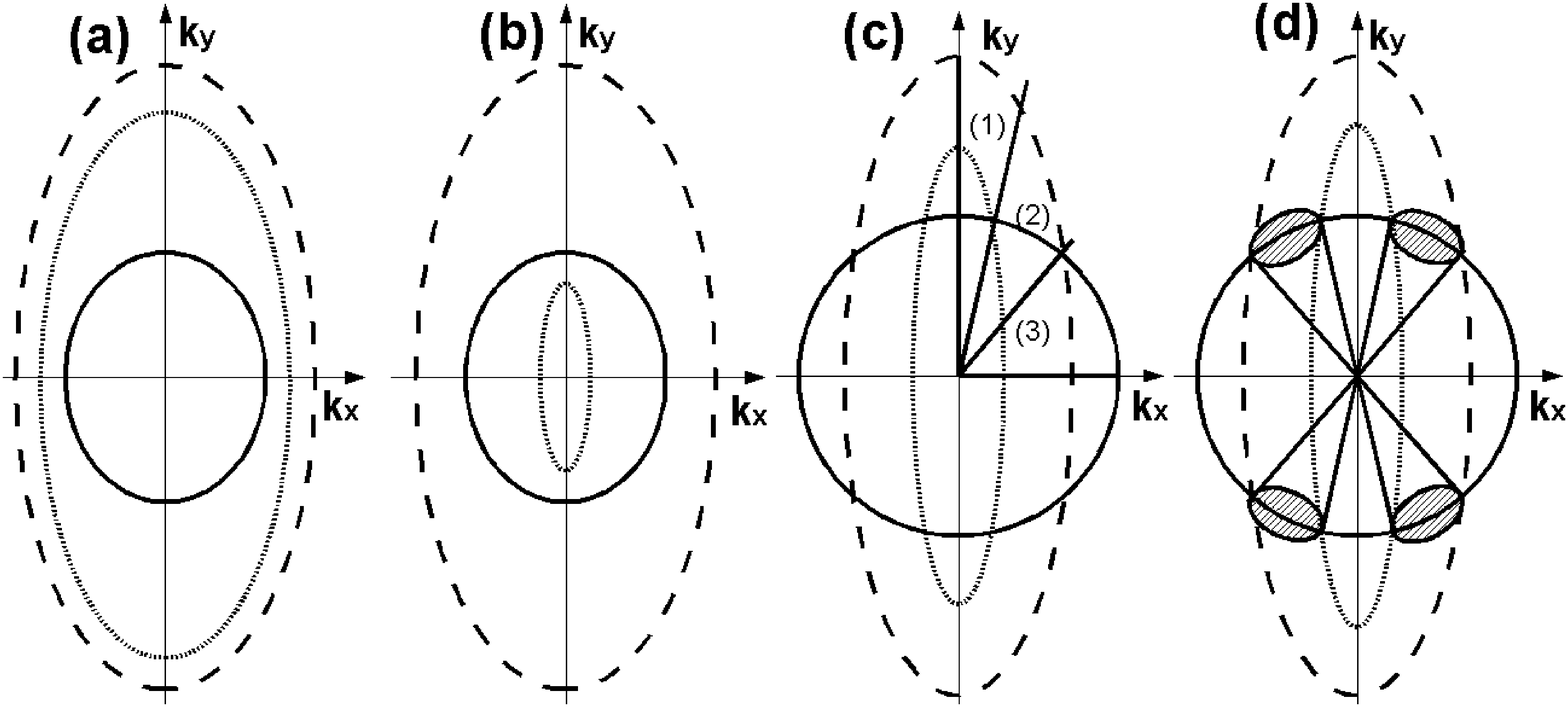}
\end{center}
\caption{2-D Fermi surfaces for superconductor $\protect\xi _{s}$ (solid
line) and for majority $\protect\xi _{\uparrow }$and minority $\protect\xi %
_{\downarrow }$ bands of ferromagnet (dashed and dotted lines
correspondingly). In (a) is depicted the situation where $\protect\xi _{s}<%
\protect\xi _{\downarrow }<\protect\xi _{\uparrow }$. In (b) is depicted the
situation where $\protect\xi _{\downarrow }<\protect\xi _{s}<\protect\xi %
_{\uparrow }$.In (c) are represented the different areas where the
integration are done. In area (1) $\protect\xi _{s}<\protect\xi _{\uparrow }<%
\protect\xi _{\downarrow }$, in area (2) $\protect\xi _{\uparrow }<\protect%
\xi _{s}<\protect\xi _{\downarrow }$ and in area (3) $\protect\xi _{\uparrow
}<\protect\xi _{\downarrow }<\protect\xi _{s}$. (d) represents the sign of $%
\ln \left( T_{c}^{0}/T_{c}^{\protect\pi }\right) $ in the different areas.
In the shaded areas, the integral is negative and in the other one, the
integral is positive. In the frontier, the integral equals zero.}
\label{FIG_toutellipse}
\end{figure}

In real compounds, the results depends on the details of the exact form of
the Fermi surfaces of superconductor and ferromagnet. Hence, the anisotropy
may have an effect on the $0-\pi $ phase transition.

To illustrate the anisotropic case, we choose elliptic Fermi surface of F
layer. The majority and minority energy spectra $\xi ^{\uparrow }\left(
\theta \right) $ and $\xi ^{\downarrow }\left( \theta \right) $ depend on
the polar angle $\theta $ in the $\left( k_{x},k_{y}\right) $ plane (see
figure \ref{FIG_toutellipse}).

In the case of strong energy band splitting between ferromagnet and
superconductor electron bands, the spectrum of majority (minority)
ferromagnet electrons can be written as $\xi ^{\uparrow \left( \downarrow
\right) }\left( \theta \right) =\xi _{s}+E^{\uparrow \left( \downarrow
\right) }\left( \theta \right) $ where $\xi _{s}$ is the superconducting
spectrum considered as isotropic one. In spite of the dependence on $\theta $%
, the situation depicted in the figures \ref{FIG_toutellipse} a) and \ref%
{FIG_toutellipse} b)\ are similar to the isotropic situation and we can
estimate the sign of $\Delta T$ without doing the integration over $\theta $%
. Thereby, the conclusions are the same as in the isotropic case.

From $\left( \ref{I_0}\right) $, we deduced that $\Delta T>0$ and the S/F/S
trilayer is in the $0-$phase if $E^{\uparrow}\left( \theta\right) $ and $%
E^{\downarrow}\left( \theta\right) $ have both the same sign for all values
of $\theta$ \textit{e.g. }$E^{\uparrow}\left( \theta\right) >0$ and $%
E^{\downarrow}\left( \theta\right) >0$ which corresponding sketch of Fermi
surfaces is presented on the figure \ref{FIG_toutellipse}a). Thus, $\Delta
T<0$ and the S/F/S trilayer is in the $\pi-$phase if $E^{\uparrow}\left(
\theta\right) $ and $E^{\downarrow}\left( \theta\right) $ have opposite sign
for all values of $\theta$ \textit{e.g. }$E^{\uparrow}\left( \theta\right)
>0 $ and $E^{\downarrow}\left( \theta\right) <0$ which corresponding sketch
of Fermi surfaces is presented on the figure \ref{FIG_toutellipse}b).
Consequently, the anisotropy of electrons band do not influence the $0-\pi$
transition phase when F layers and S layers Fermi surfaces do not intersect.

Let us consider now the case when the F layers and S layers Fermi surfaces
intersect (see figure \ref{FIG_toutellipse} c)). In this case, the
intersections delimit the regions where $\Delta T$ is positive or negative.
In order to estimate the resulting sign of $\Delta T$ without integrate over 
$\theta $, we divide the Fermi surface in three regions. And then, according
to the total surface of "positive" and "negative" region, one may conclude
which state is realized.

In the regions (1) and (3), both $E^{\uparrow }\left( \theta \right) $ and $%
E^{\downarrow }\left( \theta \right) $ have the same sign (positive in (1)
and negative in (3)). Consequently, $\Delta T$ is positive and the S/F/S
junctions exhibit a $0-$phase behavior in these regions. In the region (2), $%
E^{\uparrow }\left( \theta \right) $ and $E^{\downarrow }\left( \theta
\right) $ have opposite sign ($E^{\uparrow }\left( \theta \right) >0$ and $%
E^{\downarrow }\left( \theta \right) <0$) so, $\Delta T$ is negative and the
S/F/S junctions exhibit a $\pi -$phase behavior in these regions.

The sign of the contributions of the each region to $\Delta T$ is depicted
in the figure \ref{FIG_toutellipse} (d). If the overall size of the regions
where $\pi-$phase is realized is greater than the overall size of the region
where the $0-$phase exists, we can conclude that the phase difference
between the two superconducting layers is $\pi$. In the contrary case, the
phase difference between the two superconducting layers is $0$. These
results implies that the anisotropy of the Fermi surfaces with intersection\
and energy band shift has an important effect on the $0-\pi$ phase
transition.

We may generalize the S/F/S hybrid system by considering an arbitrary number
of F layers between the two S layers. In \cite{Andersen_2006}, the authors
found that the $0-\pi $ phase transition of
superconductor-antiferromagnet-superconductor (S-AF-S) junctions manifests a
dependence on the number of magnetic atomic layer. The AF interlayer is
composed by an even or odd number of monoatomic F layer where each F
adjacent layer have an opposite magnetization. At low temperatures the
junction S-AF-S is either a $0$ or $\pi $ junction depending on wether the
AF interlayer consists of an even or odd number of atomic layers. In this
paper, we show that in fact the $0-\pi $ phase transition depends only on
the number of ferromagnetic layers between the two S layers but not on the
relative orientation of the magnetization between adjacent F layers. We
generalize our model by adding $N-2$ F layers between the two S layers with $%
n=0$ and $n=N$. For simplicity, we consider that the ferromagnet spectrum
can be written as $\xi _{\uparrow }=E^{\uparrow }$ and $\xi _{\downarrow
}=E^{\downarrow }$ where $E^{\uparrow }$ and $E^{\downarrow }$ have opposite
sign and $E^{\uparrow \left( \downarrow \right) }\gg T_{c0}$. Thus in the
limit $t\ll E^{\uparrow \left( \downarrow \right) }$ we can express the
normal Green function between the layer $0$ and the $N$th layer like : 
\begin{equation*}
G_{0\uparrow \uparrow }^{\omega }(0,N)=\tfrac{t}{\left( i\omega -\xi
_{s}\right) ^{2}}\tprod\limits_{n=1}^{N-1}\tfrac{t}{E_{n}^{\uparrow }}.
\end{equation*}%
The critical temperature difference following the formular similar to $%
\left( \ref{autoconsistence 3}\right) $ is :%
\begin{equation}
\tfrac{\Delta T}{T_{c}^{0}}=\tfrac{7t^{2}\zeta \left( 3\right) }{8\pi
^{2}T_{c0}^{2}}\left\{ \tprod\limits_{n=1}^{N-1}\tfrac{t^{2}}{%
E_{n}^{\uparrow }E_{n}^{\downarrow }}\right\} .
\label{self_consistency_modifiee_a_N}
\end{equation}%
As it follows from $\left( \ref{self_consistency_modifiee_a_N}\right) $ the
insertion of each supplementary F layers leads to additional phase shift of $%
\pi $. As a consequence, the resulting phase difference ($0$ or $\pi $) is
determined by the number of ferromagnetic layers between the
superconductors. If the number of ferromagnetic layers is even the total
phase difference $\varphi _{t}$ induced by the ferromagnetic layers is $%
\varphi _{t}=2n\pi $ that correspond to a $0-$phase shift. For the odd
numbers of ferromagnetic layers the total phase difference is $\varphi
_{t}=\left( 2n+1\right) \pi $ \ that means the $\pi -$phase shift. Hence,
the magnetization of adjacent ferromagnetic spacers have no influence on the
fact that the system is in 0 or -phase in contrast with the result in \cite%
{Andersen_2006}.

\textbf{4. Spin-switch effect in F/S/F system} - In this section, we study
the possible inversion of the SSE in a superconductor layer sandwiched by
two ferromagnetic adjacent layers in the F/S/F structure (see figure \ref%
{FIG_SFSFSF}b)). We label $n=0,2$ so $\xi _{\sigma ,0}$ and $\xi _{\sigma
,2} $ are the electron spectra for the F layers\ and $n=1$ so $\xi
_{s,1}=\xi _{s}$ for the S layer. The BCS Hamiltonian $H_{BCS}$ and the
kinetic energy Hamiltonian $H_{0}$ are described by $\left( \ref%
{hamiltonien_general}\right) $. In \cite{A_Singh_2007}, the authors observed
the inversion of the SSE (they observed that the critical temperature in the
parallel state is higher than in the antiparallel one, $T_{cP}>T_{cAP}$) in $%
\left[ Co/Pt\right] /Nb/\left[ Co/Pt\right] $ structure and proposed an
explanation for the inversion of the SSE. In the parallel (P) state the
spin-polarized carriers can migrate from one F layer across the intermediate
S layer to the other F layer whereas in the antiparallel (AP) state this
effect is reduced and the spin polarized carriers is reflected back into the
S layer. This means that the transparency of the S-F interfaces depends on
the electron spin orientation. In our model, that implies that the transfer
energy between two layers may depend on the electron spin orientation so in $%
\left( \ref{hamiltonien_general}\right) $, instead of $t$ one, we introduce
two different energy transfer $t_{\uparrow }$ and $t_{\downarrow }$ with $%
t_{\uparrow }\neq t_{\downarrow }$. Near $T_{c}$, in the isotropic case, the
critical temperature equation is \cite{b.abrikosov_gorkov} : 
\begin{equation}
\ln \left( \tfrac{T_{cP}}{T_{cAP}}\right) =2T_{cP}\dsum\limits_{\omega
>0}\dint \func{Re}\left( \widetilde{F}_{\downarrow \uparrow ,P}-\widetilde{F}%
_{\downarrow \uparrow ,AP}\right) d\xi _{s},  \label{autoconsistence 1}
\end{equation}%
where $T_{cP(AP)}$ is the superconductor critical temperature in the P (AP)
case. Using the Green function formalism and the equations $\left( \ref%
{equations_generales}\right) $ modified with the new tunneling Hamiltonian $%
H_{t}^{P\left( AP\right) }$, we calculate the anomalous Green function in
the S layer in the P state $F_{\downarrow \uparrow ,P}^{+}\left( 1,1\right) $
and in the AP state $F_{\downarrow \uparrow ,AP}^{+}\left( 1,1\right) $.
Expanding this function in power of $t_{\uparrow }$ and $t_{\downarrow }$,
and keeping the terms proportional to $t^{4}$, we obtain the difference of
the anomalous Green function $F_{\downarrow \uparrow ,P}^{+}-F_{\downarrow
\uparrow ,AP}^{+}$ : 
\begin{equation*}
\begin{array}{c}
\widetilde{F}_{\downarrow \uparrow ,P}-\widetilde{F}_{\downarrow \uparrow
,AP}=\frac{1}{\omega ^{2}+\xi ^{2}}\left[ \tfrac{2}{\left( i\omega +\xi
\right) ^{2}}\left[ \frac{t_{\uparrow }^{2}}{i\omega +\xi _{\uparrow }}-%
\frac{t_{\downarrow }^{2}}{i\omega +\xi _{\downarrow }}\right] ^{2}\right]
\\ 
-\frac{1}{\left( \omega ^{2}+\xi ^{2}\right) }\left[ \frac{t_{\uparrow }^{4}%
}{\left( \omega ^{2}+\xi _{\uparrow }^{2}\right) }+\frac{t_{\downarrow }^{4}%
}{\left( \omega ^{2}+\xi _{\downarrow }^{2}\right) }+\tfrac{2t_{\uparrow
}^{2}t_{\downarrow }^{2}}{\left( \omega ^{2}+\xi ^{2}\right) \left( i\omega
-\xi _{\uparrow }\right) \left( i\omega +\xi _{\downarrow }\right) }\right] .%
\end{array}%
\end{equation*}%
We choose for simplicity, the energy spectra for the P case like $\xi
_{\uparrow }\equiv \xi _{\uparrow ,0}=\xi _{\uparrow ,2}$ and $\xi
_{\downarrow }\equiv \xi _{\downarrow ,0}=\xi _{\downarrow ,2}$, and for the
AP case, $\xi _{\uparrow }\equiv \xi _{\uparrow ,0}=\xi _{\downarrow ,2}$
and $\xi _{\downarrow }\equiv \xi _{\downarrow ,0}=\xi _{\uparrow ,2}$. The
sign of the right hand part in $\left( \ref{autoconsistence 1}\right) $
determines whether we deal with normal or inverse SSE. In particular, if $%
\ln \left( T_{cP}/T_{cAP}\right) $ is positive then $T_{cP}>T_{cAP}$ and so
the SSE is reversed. 
\begin{figure}[tbp]
\begin{center}
\includegraphics[scale=0.2,angle=0]{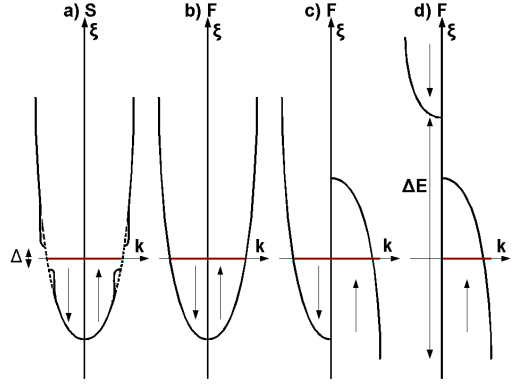}
\end{center}
\caption{(a) Spectrum of a superconducting metal (b) Spectrum of a normal
metal where $\protect\xi _{\uparrow }=\protect\xi _{\downarrow }=\protect\xi %
_{s}$. (c) Spectrum of a strong ferromagnet with reversed spectrum where $%
\protect\xi _{\uparrow }=-\protect\xi _{s}$ and $\protect\xi _{\downarrow }=%
\protect\xi _{s}$. (d) Spectrum of a strong ferromagnet with a reversed
spectrum where $\protect\xi _{\uparrow }=-\protect\xi _{s}$ and $\protect\xi %
_{\downarrow }=\Delta E\gg k_{B}T_{c}$.}
\label{FIG_graphique_spectreferroma}
\end{figure}

To verify the hypothesis of \cite{A_Singh_2007} , we consider first the
spectra $\xi _{\uparrow }=\xi _{\downarrow }=\xi _{s}$, see figure \ref%
{FIG_graphique_spectreferroma} b) and we find :%
\begin{equation}
\tfrac{\Delta T}{T_{cP}}=-\tfrac{31}{64}\tfrac{\zeta \left( 5\right) }{\pi
^{4}T_{cp}^{4}}\left( t_{\downarrow }^{2}-t_{\uparrow }^{2}\right) ^{2}
\label{transdiffspecnorm}
\end{equation}%
where we have introduced for this section $\Delta T=T_{cP}-T_{cAP}$ and $%
\zeta \left( 5\right) =1.03$. We see that the right hand side of equation $%
\left( \ref{transdiffspecnorm}\right) $ is always negative. It is the normal
SSE. We choose another spectra $\xi _{\uparrow }=\xi _{s}$ and $\xi
_{\downarrow }=\Delta E$ with $\Delta E\gg T_{c}$, then the critical
temperature difference becomes :%
\begin{equation}
\tfrac{\Delta T}{T_{cP}}=-\tfrac{1}{64}\tfrac{\left[ 31\zeta \left( 5\right)
t_{\uparrow }^{4}\Delta E^{2}+112t_{\downarrow }^{4}\zeta \left( 3\right)
\pi ^{2}T_{cP}^{2}\right] }{\Delta E^{2}\pi ^{4}T_{cp}^{4}}.
\label{transdiffspecnormdec}
\end{equation}%
The right hand side of $\left( \ref{transdiffspecnormdec}\right) $ is always
negative so $T_{cP}<T_{cAP}$. Consequently, the superconductor critical
temperature in the P case is always lower than in the AP case, $%
T_{cP}<T_{cAP}$. It is the normal SSE.

Thus, the hypothesis on the difference between $t_{\uparrow }$ and $%
t_{\downarrow }$ cannot explain the inversion of the SSE. Nevertheless, in
strong ferromagnetic, the inversion of the electron spectrum provides
another possibility to explain this unusual behavior.

Indeed, in the case of strong ferromagnet, the band splitting can be so
important that one of the electron band (here the spin up band) becomes a
hole-like (see figure \ref{FIG_graphique_spectreferroma} c)). Here, the
exchange field triggers the hole-like energy band for spin up electrons but
do not appear explicitly in the calculation. That's allows \ us to emphasize
the important influence of inverse spectra on $T_{c}$. In the very special
case where the electron spectra can be written as $\xi _{\uparrow }=-\xi
_{s} $ and $\xi _{\downarrow }=\xi _{s}$, we find :%
\begin{equation}
\tfrac{\Delta T}{T_{cP}}=\tfrac{31}{128}\tfrac{\zeta \left( 5\right) }{\pi
^{4}T_{cp}^{4}}\left( 3t_{\uparrow }^{4}-2t_{\downarrow }^{4}-4t_{\downarrow
}^{2}t_{\uparrow }^{2}\right) .  \label{transdiffspecinv1}
\end{equation}%
\ If we put $t_{\uparrow \left( \downarrow \right) }=t_{0}\pm \Delta t$ then 
$\Delta T$ is positive $\left( T_{cP}>T_{cAP}\right) $ and the spin-switch
effect is reversed if $\Delta t\gtrsim 0.1t_{0}$ (see figure \ref%
{FIG_graphique}(a)). We notice that the spin-switch effect is reversed if $%
t_{\downarrow }=0$ and normal if $t_{\uparrow }=t_{\downarrow }$. Hence, we
verify that the condition $t_{\uparrow }\neq t_{\downarrow }$ cannot explain
alone the inversion of SSE and have to be associated with the inversion of
majority electron energy band.

The minority electron band can be shifted considerably compared to the
superconductor spectrum (see figure \ref{FIG_graphique_spectreferroma} d)).
In this case, the spectra are $\xi _{\uparrow }=-\xi _{s}$ and $\xi
_{\downarrow }=\Delta E\gg T_{c}$. This situation gives :%
\begin{equation}
\tfrac{\Delta T}{T_{cP}}=\tfrac{93}{128}\tfrac{\zeta \left( 5\right) }{\pi
^{4}T_{cp}^{4}}t_{\uparrow }^{4}-\tfrac{7\zeta \left( 3\right) }{4\left(
\Delta E\right) ^{2}\pi ^{2}T_{cP}^{2}}t_{\downarrow }^{4}
\label{transdiffspecinv2}
\end{equation}%
Therefore $\Delta T$ is positive $\left( T_{cP}>T_{cAP}\right) $ for high $%
\Delta E$ and any relation between $t_{\uparrow }$, $t_{\downarrow }$ and
the opposite situation realizes if $\Delta E$ is small (see figure \ref%
{FIG_graphique}(b)). In the situation of high $\Delta E$, the inversion of
SSE is related to the inversion of spectrum of majority electrons. We see
that strong energy shifts $\Delta E$ between F and S layers favored the
inversion of the SSE.

We perform numerical calculation of $\Delta T$ in the case of spectra like $%
\xi _{\uparrow }=-\alpha .\xi _{s}+\widetilde{\Delta E}$ and $\xi
_{\downarrow }=\Delta E\gg T_{c}$ where $\alpha $ is a positive parameter
which represents the difference of effective mass between majority and
minority electron and $\widetilde{\Delta E}$ is energy shift between
electrons up and the superconductor spectrum. For simplicity, we consider $%
\widetilde{\Delta E}\simeq T_{c}$ and $t_{\uparrow }=t_{\downarrow }=t$. 
\begin{figure}[tbp]
\begin{center}
\includegraphics[scale=0.65,angle=0]{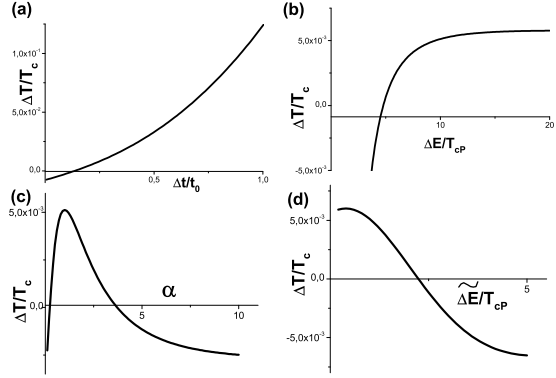}
\end{center}
\caption{(a) $\Delta T/T_{c}$ as a function of $\Delta t/t_{0}$ is presented
in the units of $\left( t_{0}/T_{c}\right) ^{4}$ in the situation $\protect%
\xi _{\uparrow }=-\protect\xi _{s}$ and $\protect\xi _{\downarrow }=\protect%
\xi _{s}$. (b) $\Delta T/T_{c}$ as a function of $\Delta E/T_{c}$ presented
in the units of $\left( t/T_{c}\right) ^{4}$ with $t_{\uparrow
}=t_{\downarrow }=t$, in the situation $\protect\xi _{\uparrow }=-\protect%
\xi _{s}$ and $\protect\xi _{\downarrow }=\Delta E\gg T_{c}$. (c) $\Delta
T/T_{c}$as a function of $\protect\alpha $ presented in the units of $\left(
t/T_{c}\right) ^{4}$with $t_{\uparrow }=t_{\downarrow }=t,\Delta E=10T_{c}$
and $\widetilde{\Delta E}=T_{c}$ in the situation $\protect\xi _{\uparrow }=-%
\protect\alpha .\protect\xi _{s}+\widetilde{\Delta E}$ and $\protect\xi %
_{\downarrow }=\Delta E\gg T_{c}$. (d) $\Delta T/T_{c}$ as a function of $%
\widetilde{\Delta E}/T_{c}$ presented in the units of $\left( t/T_{c}\right)
^{4}$with $\Delta E=10T_{c},\ t_{\uparrow }=t_{\downarrow }=t$ and $\protect%
\alpha =1$ in the situation $\protect\xi _{\uparrow }=-\protect\alpha .%
\protect\xi _{s}+\widetilde{\Delta E}$ and $\protect\xi _{\downarrow
}=\Delta E\gg T_{c}$.}
\label{FIG_graphique}
\end{figure}
For important or small difference of effective mass, there is no inversion
of SSE (see figure \ref{FIG_graphique}(c)). The inverse SSE is favored when $%
\alpha \approx 1$. Hence, the inversion of the SSE is more efficient if the
effective mass are close.

The dependence on $\widetilde{\Delta E}$ \ (see figure \ref{FIG_graphique}%
(d)) shows that the inverse SSE disappears if $\widetilde{\Delta E}\gtrsim
2T_{c}$. Thus, if the energy shift on the reversed electron energy band is
too important then the inverse SSE disappears. This condition is quite
restrictive for the appearance of the inversion of the SSE.

\textbf{5. Conclusions}-\textbf{\ } We have demonstrated an important
influence of the spectrum anisotropy of F layers and energy shift between F
and S energy spectra on the F/S hetero-structures properties.

For the S/F/S sandwiches, we have shown that the $\pi-$phase cannot appear
if the energy shift between S and F spectra is too important. Moreover, on
the example of elliptic Fermi surfaces, we show that the anisotropy of F
layer spectra may provide rather restrictive conditions of existence of the $%
\pi -$phase. We have also considered a more general S/F/S structure by
including several F layers between the two S layers. The $0-\pi$ phase
transition is influenced by the even or odd number of F layer independently
of the relative magnetization between adjacent F layer.

For the F/S/F trilayers, we propose a possible explanation of the previously
observed inverse SSE. In strong ferromagnet, the high energy band splitting
may imply an inversion of one of the electron energy band. This type of
electron spectrum in F layer could explain the anomalous SSE. We also study
the influence of the effective mass difference and the energy shift on the
inverse spectra.

\textbf{Acknowledgements:} The authors thank J. Cayssol for useful
discussions and helpful comments. This work was supported by the French ANR
Grant No. ANR-07-NANO-011:ELEC-EPR.


\begin{thebibliography}{99}
\bibitem{Efetov_2005} \Name{Bergeret F,Volkov A. \and Efetov K.} %
\REVIEW{Rev. Mod. Phys.}{77}{2005}{1321}.

\bibitem{Golubov_2006} \Name{Golubov A., Kupriyanov M. \and II'ichev E.} %
\REVIEW{Rev. Mod. Phys.}{76}{2004}{411}.

\bibitem{buzdin(2005)} \Name{Buzdin A.I.} 
\REVIEW{Rev. Mod.
Phys.}{77}{2005}{935}.

\bibitem{Yamashita_2005} \Name{Yamashita T. \emph{et~al}} 
\REVIEW{Phys.
Rev. Lett.}{95}{2005}{097001}.

\bibitem{Bulaevskii_19822} 
\Name{Buzdin A.I., Bilaevskii L.N. \and Panjukov
S.V.} \REVIEW{JETP Lett.}{35}{1982}{178}.

\bibitem{Ryazanov_Aarts_2001} \Name{Ryazanov V. \emph{et~al.}} 
\REVIEW{Phys.
Rev. Lett.}{86}{2001}{2427}.

\bibitem{Kontos_2002} \Name{Kontos T. \emph{et~al.}} 
\REVIEW{Phys. Rev.
Lett.}{89}{2002}{137007}.

\bibitem{Andreev_Buzdin_OsgoodIII_1991} 
\Name{Andreev A.V., Buzdin A.I.
\and Osgood III R.M.} \REVIEW{Phys. Rev. B}{77}{1991}{10124}.

\bibitem{houzet_buzdin.2001_2} \Name{Houzet M., Buzdin A. \and Kulic M.} %
\REVIEW{Phys. Rev. B}{64}{2001}{184501}.

\bibitem{Mac_Laughlin_1999} \Name{McLaughlin A.C. \emph{et~al.}} %
\REVIEW{Phys. Rev. B}{60}{1999}{7512}.

\bibitem{De_Gennes_1966} \Name{De~Gennes P.G.} 
\REVIEW{Phys.
Lett.}{83}{1966}{10}.

\bibitem{Buzdin_vedyayev_1999} 
\Name{Buzdin A.I., Vedyayev A.V. \and
Ryzhanova N.V.} \REVIEW{Europhys. Lett.}{48}{1999}{686}.

\bibitem{Tagirov_1999} \Name{Tagirov L.R.} 
\REVIEW{Phys. Rev.
Lett.}{83}{1999}{2058}.

\bibitem{Deutscher_1969} \Name{Deutscher G., Meunier F.} 
\REVIEW{Phys. Rev.
Lett.}{22}{1969}{395}.

\bibitem{Hauser_1969} \Name{Hauser J.J.} 
\REVIEW{Phys. Rev.
Lett}{23}{1969}{374}.

\bibitem{JY_Gu_2002} \Name{Gu J.Y. \emph{et~al.}} 
\REVIEW{Phys. Rev.
Lett}{89}{2002}{267001}.

\bibitem{Aarts_J_2004} \Name{Rusanov A.Y. \emph{et~al.}} 
\REVIEW{Phys. Rev.
Lett}{93}{2004}{057002}.

\bibitem{O_Birge_2006_1} \Name{Birge N.O., Pratt W.P. \and Moraru I.C.} %
\REVIEW{Phys. Rev. Lett}{96}{2006}{037004}.

\bibitem{Miao_2007} \Name{Miao G.X. \emph{et~al.}} 
\REVIEW{Phys. Rev.
Lett}{98}{2007}{276001}.

\bibitem{Aarts_J_2006} \Name{Rusanov A.Y., Habraken M. \and Aarts J.} %
\REVIEW{Phys. Rev. B}{73}{2006}{060505}.

\bibitem{O_Birge_2006_2} \Name{Birge N.O., Pratt W.P. \and Moraru I.C.} %
\REVIEW{Phys. Rev. B}{74}{2006}{220507(R)}.

\bibitem{Steiner_2006} \Name{Steiner R., Ziemann P.} 
\REVIEW{Phys. Rev.
B}{74}{2006}{094504}.

\bibitem{Stamopoulos_2007} \Name{Stamopoulos D., Manios E. \and Pissas M.} %
\REVIEW{Phys. Rev. B}{75}{2007}{014501}.

\bibitem{A_Singh_2007} 
\Name{Singh A., S\"{u}rgers C. \and L\"{o}hneysen
H.v.} \REVIEW{Phys. Rev. B}{75}{2007}{024513}.

\bibitem{Santamaria_2008} \Name{Santamaria J. \emph{et~al.}} 
\REVIEW{Phys.
Rev. B}{78}{2008}{094515}.

\bibitem{Zhu_2009} \Name{Zhu J.\emph{et~al.}} %
\REVIEW{arXiv}{}{2009}{0903:0044}.

\bibitem{Nachtrab_2006} \Name{Nachtrab T. \emph{et~al.}} 
\REVIEW{C.R.
Physique}{7}{2006}{69}.

\bibitem{b.abrikosov_gorkov} 
\Name{Abrikosov A.A., Gorkov L.P.  \and
Dzyaloshinsky I.} 
\Book{Methods of quantum fiel theory in statistical
physics} \Editor{Prentice Hall} \Year{1963}.

\bibitem{Andersen_2006} \Name{Andersen B.M. \emph{et~al.}} 
\REVIEW{Phys.
Rev. Lett.}{96}{2006}{117005}.
\end{thebibliography}
\end{document}